D. V. Sarafopoulos

2020

# The dynamo action for red dwarfs and red giant and supergiant stars

D. V. Sarafopoulos

2020

# The dynamo action for red dwarfs and red giant and supergiant stars

D. V. Sarafopoulos

Department of Electrical and Computer Engineering,

Democritus University of Thrace, Xanthi, Greece

sarafo@ee.duth.gr

2020

# The dynamo action for red dwarfs and red giant and supergiant stars

Abstract

# Abstract

We investigate the possibility to apply the already suggested by Sarafopoulos (2017, 2019) main concept of dynamo action to red dwarfs and red giant and supergiant stars. Thus, we attempt to establish a unified dynamo action, being potentially at work at widely varying stellar domains. Thus, the powerful, unique and leading entity generating the primary stellar magnetic field remains the so-called "Torus structure". ***Within the Torus the same sign charges are mutually attracted and the Torus could be simulated as a superconductor.*** An existing gradient of the rotation rate accumulates net charge in the Torus, and the resulting toroidal current becomes the driving source of the magnetic field. In turn, there is a complicated network of secondary interactions that affect and modulate the whole star's magnetic behaviour. Our dynamo action is potentially at work in fully and partly convecting stars. ***A major finding is that the strength of the magnetic field, for stars of the same spectral type, is essentially controlled by the size of the star;*** we suggest that, the larger the star, the deeper inward the Torus is formed. The formation of a single Torus is essentially associated with a large-scale strong, poloidal and axisymmetric magnetic field topology, whereas the generation of a weaker multipolar, non-axisymmetric field configuration in rapid evolution can result from a double-Torus structure (like the solar case). The same basic concept is scaled up or down; in parallel, we conclude that the tachocline does not play the supposed role in the classical αΩ solar dynamo action. Moreover, *we identify four key parameters* associated with the magnetic field of red dwarfs, giants and supergiants: **First**, the rotation speed; **second**, the steepness of the radial gradient of the rotation rate ($\nabla\Omega$) in the shear layer; **third**, the distance of the Torus from the photosphere and **fourth**, the cross-sectional area of the Torus. The third and fourth key parameters are introduced for the first time. In our model, the observation that in the longest rotation periods there are active stars exclusively within the category of late M dwarfs is interpreted based on the extremely high gradient of rotation rate (in this category) that surpasses the role of the minimized rotation speed.

# 1. Introduction

Our approach, aiming to further establish a dynamo action beyond the conventional model, is mainly and inevitably qualitative, at this phase. With the present (third) work, this author attempts to integrate his effort producing a unified stellar dynamo model based equally on electric currents and magnetic fields. We exclusively use the equations of Maxwell and the Lorentz force, while certainly additional stellar properties, like the differential rotation, radial density discontinuities, rotational speeds, etc. are all taken into account, too. Every MHD approach is dismissed.

All the stars considered in this work essentially belong to the same spectral type; however, a part of them, like the M4-M7 type stars, belongs to the fully convective sub-category, while the early M-type stars belong to the class with convective envelopes; moreover, a third class includes the red giants and red supergiants. We basically move vertically in the Hertzsprung-Russell diagram and, consequently, the ultimate concept related to our suggested dynamo action is scaled up and down and tested in different stellar domains. Thus, in this work a lot of stars are engaged with extremely varying parameters like the magnitude of the magnetic field and its main configuration, the size of the stars, the plasma densities, the degree of ionization, the depth in which the magnetic field is mainly generated, the distribution of starspots on the surface, the level of the magnetic activity, the existence of repetitive activity cycles, the rotation periods, the metallicity, the internal structure, etc. It could be particularly significant if all these variable parameters and material properties cooperate and finally obey to the same ultimate principle dictating the basic magnetic behaviour for all these stars. In our first work (Sarafopoulos, 2017), an entirely new solar dynamo action was suggested based on the so-called powerful "Torus structure" with its exotic properties. In our second work (Sarafopoulos, 2019), the same dynamo action was extended to bigger Main Sequence stars, like those belonging to the A, Ap and Am classes; in parallel, ***the initial solar model was further refined and the role of the differential rotation was precisely explained***. In this work, the same dynamo action is applied on red stars, with much smaller and much larger radii (as compared to those studied in the first two works), on or off the Main Sequence. Our model, at this stage, being potentially at work on various environments, further improves a few aspects related to the initially proposed solar dynamo action. In general, in a highly complicated system of huge size, like a star, every researcher faces two main obstacles: First, it is extremely difficult to really identify the fundamental involved parameters and second, it is equally hard, almost impossible, to comprehend all the numerous secondary mechanisms leading to diverse responses.

In every astronomy or astrophysics book, you can read that the magnetic fields in the Sun and all the partially convective stars (e.g., all the G-K type stars and the early-M dwarfs) are produced by the so-called αΩ dynamo mechanism, operating at the interface between the convective envelope and the radiative core. The αΩ dynamo is hypothetically maintained by the combined action of differential rotation (Ω-effect) and cyclonic convection (α-effect). The action of differential rotation on a weak poloidal field, at the base of the convective zone, results in a large-scale and predominantly toroidal subsurface field (Parker 1975). However, when this toroidal field emerges through the surface of the Sun, the magnetic flux tubes are oriented in a primarily radial direction in the photosphere. In fully convective stars (e.g., stars with mass less than ~0.35 $M_\odot$), the lack of radiative core is expected to preclude the αΩ dynamo; in this case, an alternative model known as the "$\alpha^2$ dynamo", which is exclusively maintained by convection, was proposed by Roberts and Stix (1972).

In a different philosophy from that involved in the αΩ and $\alpha^2$ dynamo models, we are distanced from every MHD approach, while aiming to determine the ultimate mechanism triggering, maintaining and potentially reversing the stellar magnetic fields. A sustained and well-defined ring-type toroidal electric current, in the star's interior, becomes the cornerstone in our suggestion: The so-called "Torus" entity is the most precious entity, the fundamental structure and the powerful engine, generating the primary stellar magnetic field. Certainly there are a lot of secondary functions and a complicated network of interactions, but the Torus is the heart, triggering, supporting and modulating the stellar overall magnetic behaviour. The Torus essentially carries a net charge; consequently, its rotational speed generates an intense azimuthal-toroidal electric current. Thus, the Torus could be simulated as a superconductor. ***Within the Torus, the existing charges (of the same sign) do not repeal each other; instead they are attracted***. Moreover, additional charges are progressively attracted inside the Torus, increasing its current and poloidal magnetic field strength, and probably widening its cross-sectional area. This exotic material property is due to the fact that the local charge velocity is greater than the phase velocity of an electromagnetic wave; in this case, ***the attractive magnetic force between two charges is greater than the repulsive electric one*** (for more details look at Sarafopoulos 2017 and 2019). Once a single or two-Torus structure, is built up, the whole star is affected by this unique and leading electromagnetic domain. In turn, a distinct sub-surface layer, with its own differential rotation, is particularly affected, modulating (in a large degree) the exo-photosphere magnetic field. Eventually, in our view, the Torus with its toroidal current dictates the processes usually attributed to the so-called $\alpha\Omega$ and $\alpha^2$ actions.

The dynamo action for red dwarfs is elaborated in section 2; separately, for the M4-M7 type stars producing dipole magnetic field and the M1-M3 stars producing weak and complex magnetic field. However, we do not support the idea that exclusively the M4-M7 (M1-M3) stars are related to dipole (more complex) field structures; conversely, later on we argue that the ultimate dynamo action is the same for both sub-categories. The difference is that the magnetic field is manifested with two radically different configurations. In section 3, we treat the dynamo action for red giants and red supergiants. Finally, in the discussion section 4, we stress our conclusion that the strength of the magnetic field for red dwarfs and red giants and supergiants is directly depended on the star's radius. Given that today there are a lot of contradictions between different dynamo actions at different stellar entities, the necessity for a unified model is imperative. In addition, for the first time, we resolve the issue related to the origin of the prevailing toroidal magnetic field component in solar-type stars. Our model is scrutinized in the light of contemporary observational results.

# 2. The dynamo action for red dwarfs

The approach methodology is as follows: Initially, two distinct and completely separated mechanisms of dynamo action are considered. The one that matches better with the early M-type stars producing weak-complex magnetic field and the other which is related to the late M-type stars producing dipole magnetic field. However, this is a preliminary approach; later on, we shall argue that the ultimate dynamo action is one and unique for both categories. As a matter of fact, the same dynamo action is manifested with two radically different configurations of the magnetic field giving the false impression about two different dynamo models. Eventually, based on these completely separated magnetic field patterns, we shall encompass the real world observations, being much more complicated.

## 2.1. The dynamo action for the portion of M4-M7 type stars producing dipole magnetic field

### 2.1.1. Radial profile of rotation rate for fully convective stars

First, for the fully convective red dwarfs with mass less than 0.35 $M_\odot$, we adopt the following macroscopic view for the star's rotation rate ($\Omega/2\pi$) profile versus its normalized radius ($r/R_*$): We assume that this profile is characterized by three particularly important and distinct shear layers related to the radial differential rotation (look at Fig. 1, upper panel- orange line, corresponding to very low latitudes). First, the so-called layer **R1**, which is characterized by a very low $\nabla(\Omega/2\pi)$ extended outward from the star's very core region (having the highest density) to an indicative $R_d$-value of ~0.76 $r/R_*$. Second, the shear layer **R2**, which is characterized by an abrupt intensification of the $\Omega/2\pi$ occurring at ~0.76-0.82 $r/R_*$. And third, the shear layer **R3** with a profound drop of $\Omega/2\pi$ that occurs in a sub-surface thin layer with an indicative fractional thickness ~0.05 $r/R_*$. The reason dictating the necessity for each layer is developed in the next paragraphs.

***Shear layer R1:*** The nominal value of $\Omega/2\pi$ corresponds to the core region, wherein thermonuclear fusion of hydrogen takes place. The helium build-up for the fully convective stars does not occur at the core, since the helium produced by the fusion of hydrogen is constantly remixed throughout the star. Mass and energy is continuously transported from the core to the surface and ***the ceaseless outward mass transportation inevitably dictates that the rotational speed gradually decreases in the region R1.*** The value of ~0.76 $r/R_*$ is an indicative one, with a fractional radius greater than that corresponding to the innermost edge of the solar tachocline at ~0.68 $r/R_\odot$; the reason is that the mean plasma density of red dwarfs is much higher. We know that, the lower the effective temperature (for a main-sequence-star) is, the higher will be the mean star's density (Zombeck, 2007). The middle panel of Fig. 1 shows the adopted typical radial profiles of the rotation rate for an early M-type star, at different latitude angles (i. e., 0°, ±15°, ±30° and ±45°). Obviously in this case, we borrow the solar radial profiles given that mass transfer does not take place and, consequently, there is not any layer R1. It is not meaningless to underline here that, in our view, the solar dynamo action is essentially triggered by the large density drop associated with the tachocline; we actually need a shear layer like the R2 one, not exclusively the boundary associated with the radiative and convective regions. The abrupt density decrease probably suffices producing the powerful Torus structure within the layer R2.

***Shear layer R2:*** Outward of the $R_d$-value (as previously defined), a radically different process dramatically changes the profile of the rotation rate. In region R2 the prominent feature is that the star's density abruptly decreases, while at the same time the mass transportation essentially terminates. Indicative radial profiles of density are given in the bottom panel of Fig. 1, for fully convective (M4-M7 stars) and early M1-M3 stars. ***In the layer R2, the rotational speed abruptly increases as the angular momentum of the outflowing plasma is redistributed.*** This process is analogous to that taking place in the solar convective region wherein exclusively angular momentum is exchanged; there is no mass transfer in R2. The layer R2 is probably associated with the meridional cells of plasma circulation.

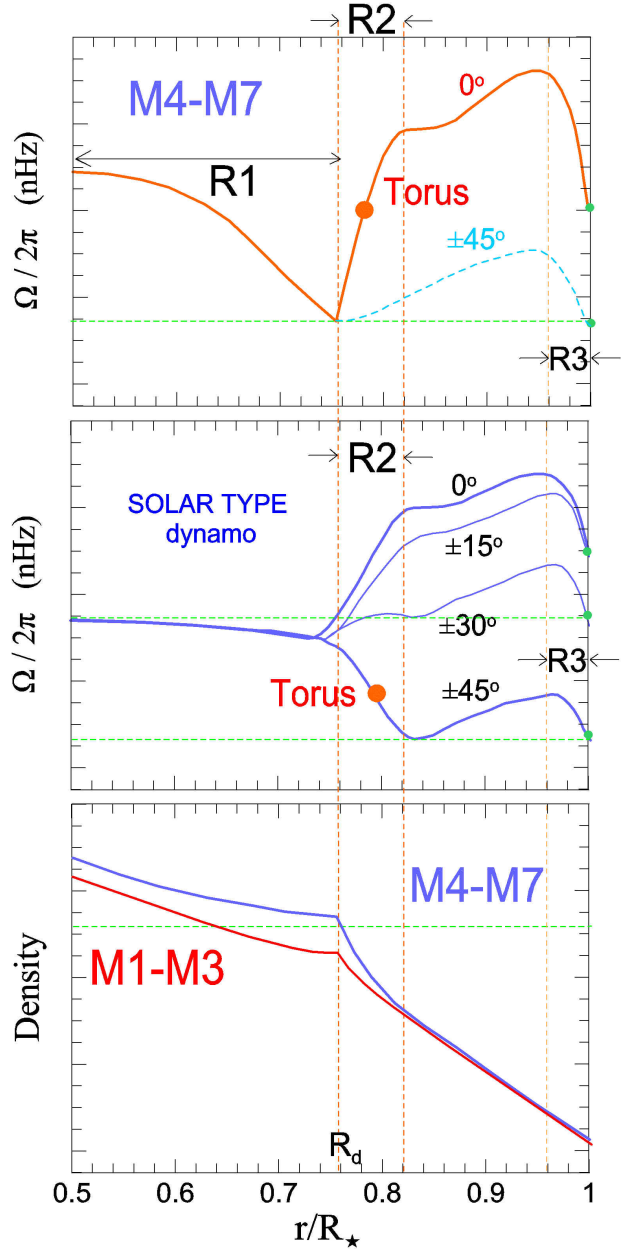

*Fig. 1. Radial profile of the rotation rate (Ω/2π) for fully convective M4-M7 type stars (first panel) and M1-M3 type stars (middle panel) associated with dipole and weak-complex magnetic field, respectively. In the first category a single Torus is formed within the shear layer R2 (and marked by an orange solid dot), whereas in the second category two Torus structures are simultaneously formed symmetrically with respect to the equatorial plane. The layer R1 results from the radial outward mass transportation, while the layer R2 is established via a process of angular momentum redistribution. The third panel shows the radial density profiles assumed for the two above stellar categories.*

***Shear layer R3:*** This layer adjacent to the photosphere (with thickness ~0.05 $R_*$) is assumed playing the same role as in the solar case. Moreover, the radial differential rotation is probably related to a latitudinal one visible on the surface. For instance, for a fully convective star, at the latitude of 45°, we would probably anticipate the radial profile shown with the light blue dashed-line in the upper panel of figure; it results as synthesis of the processes taking place in layers R1 and R2, plus the latitudinal differential rotation. Consequently, one reasonably infers that the fully convective star should reveal a surface differential rotation; however, the latter is not consistent with existing observations that a fully convective red dwarf essentially rotates as a solid body (Donati et al., 2008; Morin et al., 2008). This contradiction is resolved in the following sub-paragraphs, discriminating between two distinct phases, the "pre-Torus age" and the "post-Torus age".

## 2.1.2 Differential rotation in the "pre-Torus age"

In Fig. 1 (top panel) the indicative rotation rates, for two latitudinal profiles (i.e., 0° and ±45°) for fully convective red dwarfs, are displayed. The latitude depended drop of the rotation rate on surface, occurring in region R3, is mainly due to the meridional plasma flow transporting angular momentum to higher latitudes. For instance, on the Sun the poleward plasma velocity is ~20 ms$^{-1}$ on the surface, and this process redistributes the angular momentum in a surface layer of thickness ~0.05 $R_\odot$. Consequently, we normally should observe differential rotation on the surface of a fully convective star, as it is shown in the upper panel; however, the situation is catalytically modified by the developed magnetic field and specifically by the powerful Torus magnetic field.

## 2.1.3. Differential rotation in the "post-Torus age".

After the Torus formation, there is a significant deviation from the previously exhibited scenario. The main change is that the surface layer does not develop differential rotation; thus, the star is modelled as a revolving solid body. According to our explanation (for a fully convective star) the latter is mainly due to the formation of a single Torus over the equatorial plane and within the shear layer R2. The Torus current produces very strong poloidal magnetic field that finally prevents the meridional circulation of plasma in the layer R3. The "exodus channel" conveying angular momentum at low latitudes is blocked and the surface layer is isolated. Consequently, without the meridional circulation, there is probably none mechanism supporting the differential rotation on the stellar surface. Moreover, it should be noted that the differential rotation is probably affected in the region inward of the layer R3, too; however, given that the Torus, being the main engine for the poloidal magnetic field, is already built up inside the star, the differential rotation is lacking its "pre-Torus" dominant role. Once the Torus is formed, the shown radial profiles will change; at least the profiles for the latitudes of $0^o$ and $45^o$ will have the same value of rotation rate on the surface.

### 2.1.4. The activity cycles

We deal with a promising repetitive activity cycle possibly taking place in a fully convective star that involves the production of a strong poloidal magnetic field. Certainly, the M4-M7 stars developing weak magnetic field do not belong to this category. The whole cycle includes the four distinct phases of birth, growth, destruction and regeneration of a single Torus. In the first place, the star inaugurates its activity cycle and forms a single Torus structure. Then, there is a monotonic growth of the Torus electric current that unavoidably leads to its saturation and degeneration. Finally, a new cycle is triggered with an oppositely directed "seed magnetic field". A brief description of the whole procedure is as follows:

*Phase 1: The formation of a single Torus*

We present this fundamental process in several distinct steps visualized in the upper part of Fig. 2:

a). ***The convection process transports mass radially and homogeneously outwards***. This is particularly intense throughout the layer termed R1 in Fig. 1, wherein the rotation rate smoothly and progressively decreases. Furthermore, given that the angular momentum is essentially dependent upon the low latitude mass (related to larger radii), then it is implied that ***the rotation rate variation, for the region R1, will have its maximum value over the equatorial plane.***

b). Beyond the region R1 the mass transfer essentially terminates; exclusively angular momentum redistribution takes place. The rotation rate in region R2 abruptly increases (and mainly at low latitudes) as a consequence of the meridional plasma circulation. Accordingly, *the outward pointing gradient of rotational speed is particularly intense at low latitudes*.

c). Furthermore, the magnetic force $\mathbf{F}= -e\mathbf{u}_e \times \mathbf{B}_o$, acting on electrons, has its highest values at lowest latitudes, too. The seed magnetic field $\mathbf{B}_o$ is initially assumed being northward directed; thus, the low latitude electrons preferentially move inward and accumulate at the shear layer R2 characterized by decreased rotational speeds $\mathbf{u}_e$. Additionally, the meridional circulation dictates that the low latitude ions move radially outward and redistribute the angular momentum. In result, the net negative charge density of region R2 has its highest value at low latitudes, where the electrons have their highest chances to coalesce forming a structure with a single Torus. The negatively charged "plasma pockets", within the Torus, are mutually attracted given that these regions are placed inward of the spheroid surface defined by $u_e=c′$. Again, we remind that if the $u_e$ is greater than $c′$, then the attractive magnetic force dominates against the repulsive electric one (Sarafopoulos 2017 and 2019).

d). The same magnetic force accelerates (the inward moving) electrons within the layers R1 and R3, wherein the rotation rates increase. As a result, these layers accumulate positive charge; the $Q_S$ within the layer R3 and the $Q_{IN}$ in the layer R1.

e). The newly formed Torus attracts more and more electrons inward and monotonically strengthens its poloidal magnetic field. The intense toroidal current, being azimuthally variable, produces an exceptional strong (and longitudinally variable) poloidal magnetic field.

f). Once the Torus is formed, its poloidal magnetic field modifies the meridional plasma flow. The outward plasma flow, at low latitudes, is prevented by the intense poloidal magnetic field; the meridional circulation breaks and its role is minimized. There is no any exodus channel transporting angular momentum outward from the layer R2. In effect, *the Torus formation finally switches off the surface differential rotation*; during this long-lasting phase of the star's activity none significant differential rotation, on the surface, is anticipated.

g). The exo-photosphere (southward directed) magnetic field is essentially a dipole-like axisymmetric field. This is synthesized from the poloidal magnetic field of Torus plus the poloidal field from the rotating sub-surface charge $Q_S$.

## Phase 2: The Torus destruction

During the Torus growth phase, its total charge $Q_T$ steadily increases and the same does the sub-surface charge $Q_S$ (of region R3), as well as the volume charge $Q_{IN}$ (of region R1). Obviously $Q_T=Q_S+Q_{IN}$. Moreover, the rotating charges, being azimuthally flowing currents, develop mutually attractive or repulsive forces; moreover, it is noted that the magnetic forces prevail against the electrostatic ones inward of the curve $u_e=c'$. Eventually, the increasing ion-charge $Q_{IN}$ will force the Torus electron-charge to move outwards. However, the motion of Torus would displace it out of the surface $u_e=c'$; in this case, it will lose its exotic property as "an electron trap". The Torus gradually degenerates and the toroidal charge $Q_T$ diffuses outward. The charge $Q_S$ automatically disperses out as soon as the poloidal magnetic field collapses. The only remaining charge is the $Q_{IN}$ that initiates the new magnetic cycle. In this cycle, the "seed magnetic field" is oppositely directed as compared to that in the preceded cycle.

## Phase 3: The next activity cycle

The positively charged layer R1 affects all the outer layers through its southward directed magnetic field; the electrons move to the surface. These electrons, in the shear region R2, are accelerated, forming a positively charged Torus. Then, the leading poloidal magnetic field of the Torus dictates the whole magnetic behaviour of the star. As a result, the sub-surface layer $Q_S$ and the $Q_{IN}$ will become negative. That is, although initially the $Q_{IN}$ was positive, then, after a transition period, it is switched to a negative value.

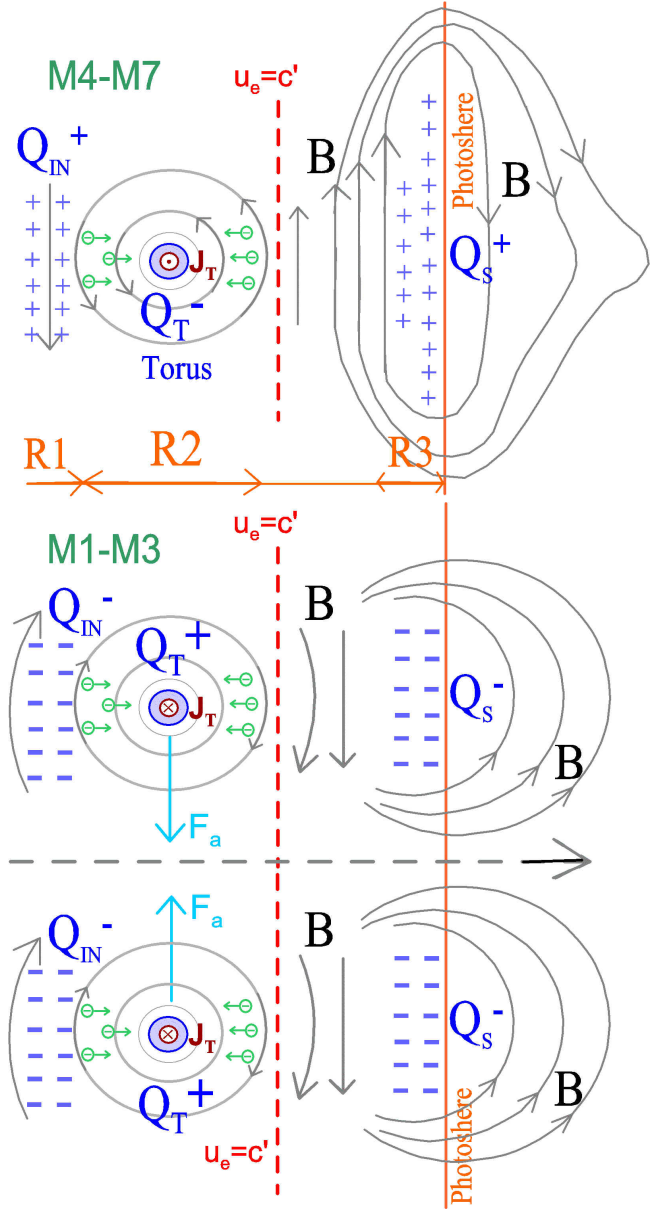

***Fig. 2***. *Schematic illustrating partially the dynamo action for the M-type stars. The seed magnetic field $B_o$ is northward directed. The surface defined by $u_e=c'$ and the photosphere are denoted with red-dashed and orange lines, respectively. The shear layers R1, R2 and R3 correspond to those of Fig. 1 and are related to the charges $Q_{IN}$, $Q_T$ and $Q_S$. In the upper part, the late M4-M7 type dynamo action is displayed associated exclusively with the formation of a single negatively charged Torus. The Torus magnetic field along with the field resulting from the sub-surface positive charge produce the intense poloidal magnetic field observed outside the photosphere. In the lower part, the dynamo action for the early M1-M3 type stars is sketched only for the case of two positively charged Torus structures.*

## 2.2. The dynamo action for the portion of M1-M3 type stars producing weak and complex magnetic field

This dynamo is in this case essentially a scaled down solar type dynamo model; the basic morphology and processes obey to the same principles. However, the approach from a different point of view enables us to slightly improve the already published solar type dynamo action (Sarafopoulos 2017, 2019). Specifically, we are further convinced that the role of tachocline is in particular not so important; the tachocline should be deprived of the properties endorsed by the widely accepted αΩ dynamo model. The solar plasma density abruptly drops just outside the radiative region and this steep density gradient is the decisive factor for the dynamo. Furthermore, the steep density gradient together with the meridional plasma circulation, establish the crucial process of angular momentum redistribution. Therefore, the rotational velocity will unavoidably increase just outside the radiative region (at low latitudes). In this way, the resulting shear layer (of ∇Ω) paves the way for the generation of the powerful two-Torus structure. In the preceded subsection 2.1, a similar density gradient is assumed existing (for the fully convective M type stars) and finally producing the strong and well organized magnetic field in this class. In result, the dominant role of the radial density gradient unifies the solar type and the non-solar type dynamos; we identify one and unique ultimate dynamo action. The creation of a single or two-Torus structure is a feature of secondary importance, although the magnetic field is morphologically radically different in each case. A single Torus is associated with a strong poloidal and largely axisymmetric magnetic field with large mean value, whereas the two-Torus structure is related to a very low mean magnetic field showing a prevailing toroidal component. At this point, it should be stressed that in our solar dynamo action, the mechanism producing the "prevailing toroidal component" was not clearly discussed in the past. Given that the subsurface charge, being largely responsible for the photosphere magnetic field, is azimuthally distributed, then a poloidal rather than toroidal magnetic field should be anticipated. We extensively scrutinize this issue and the specific mechanism at work in the discussion section. Although it is stressed that we presently study the situation of a M1-M3 dynamo action related to a solar type one; however, we briefly underline a few fundamental steps, which are associated with the lower panel of Fig. 2:

*Phase 1: The formation of two-Torus structure*

We elaborate this phase in order to clarify and further stress the differences from the preceded category of late M-type stars. Main processes are as follows:

a). The Lorentz force $\mathbf{F} = -e\mathbf{u}_e \times \mathbf{B}_{seed}$ steadily moves electrons inward; the seed magnetic field $\mathbf{B}_{seed}$ is northward directed as in the previous subsection. We adopt an ($\Omega/2\pi$, $r/R_*$) diagram like that included in the middle panel of Fig. 1; it is essentially similar to that constructed for the Sun (e.g., Kosovichev et al., 1997; Schou et al., 1998). This (scaled down) diagram displays the differential rotation occurring in the outer convective region for four contours corresponding to 0º, 15º, 30º and 45º latitudes.

b). As the moving electrons cross the shear layer R2, at latitudes of about ±45º, they are subject to an intense acceleration. The reason is that, at that place, there is an inward directed strong gradient for the rotational speed. Consequently, within this layer, positioned inward of the curve $u_e=c'$, a region with net positive charge $Q_T$ can be formed and progressively evolved into the Torus structure. Inward of the layer R2, the electrons accumulate since the plasma density becomes much higher; thus, a negatively charged region $Q_{IN}$ is formed. The currents related to $Q_T$ and $Q_{IN}$ are mutually repulsive; in contrast, the two Torus currents are mutually attractive. And the two-Torus structure is the crucial entity dictating the whole magnetic activity. At lower latitudes the electrons fail to form a viable Torus.

c). The Torus charge gives birth to an eastward directed current with its own poloidal magnetic field. Furthermore, this poloidal field, expelling additional electrons outwards, progressively increases the positive charge of the Torus; there is a positive feedback effect.

d). It is noted that in the region extended all the way from the Torus to the photosphere, the electrons are forced by the Torus poloidal magnetic field and steadily move outwards. In this way, an "electron channel" is formed going through the shear layer R3 characterized by an inward directed gradient of rotation speed. Eventually, a negatively charged subsurface layer is formed due to the local electron deceleration. This charge $Q_S$ produces a westward current generating the sunspot pairs on the star's surface and the huge magnetic loops emerging out of the photosphere.

*Phase 2: The Torus system destruction*

Each one newly formed Torus is subject to an asymmetric equatorward growth due to the (equatorward pointing) gradient of rotation speed. Moreover, the two Torus structures are mutually attracted by their own parallel toroidal currents. Their merging is inevitable, leading to an abruptly enhanced toroidal current and poloidal magnetic field alike. Then **the single Torus is steadily extended inward,** while increasing its cross-sectional area; finally the Torus is engulfed inside the high density region termed R2. The destruction of the Torus will directly annihilate the amount of the subsurface charge $Q_S$. The only remaining charge in the system will be a portion of the positive Torus charge in region R2.

*Phase 3: The next activity cycle*

At this stage the positively charged central region of the star will initiate the next activity cycle. The seed field is reversed to the southward direction. The electrons will move outward and decelerate in the layer R2; two negatively charged Torus structures will be formed, while the sub-surface charge will be positive, in region R3.

## 2.3. The real dynamo action for all the M1-M7 stars

Observationally a portion of the M4-M7 type stars shows weak and complex magnetic field. Can we incorporate such a result in our model? Really, a fully convective star lacking the region L1 will probably develop a solar type dynamo. For instance, if the star is a newborn one, then it would be realistic to assume that L1 is missing.

Moreover, there are stars in the M1-M3 category displaying strong and poloidal magnetic field. Can we accommodate this observation in our model? In this situation, we pay attention to two factors: First, there are stars where the convection region is extremely enhanced. The star does not involve only the ~2% of the stellar mass, as it happens with the Sun, but a much higher percentage of mass. Thus, the convection process may suffice to redistribute the angular momentum and create the region L1. Second, an old M1-M3 type star may probably have more chances to produce the region L1. The Helium production is equivalent to inward mass transfer; thus, the radial rotation rate steadily drops in the convection region.

# 3. The dynamo action for red giants and supergiants

Throughout the preceded analysis, it is clearly suggested that ***a star with steep radial density variation and/or an abrupt rotational rate change can generate magnetic field***; the required condition is that this change must occur inward of the curve satisfying the condition $u_e=c′$. Presumably, the powerful dynamo mechanism is based on the so-called "Torus structure". Thus, we infer that the red giants and supergiants, with probably magnetized progenitors, preserve their magnetic behaviour throughout their whole evolutionary track. The Torus, once formed in the star's interior, would be probably conserved during successive expansions and/or contractions occurring in the outer layers. We know that each core contraction due to gravo-thermal catastrophe will expand the star; in contrast, each time the core is forced to expand, the star will contract. Thus, one may plausibly anticipate that all these possible phase transitions, occurring as the star evolves leaving the Main Sequence, are tightly related to the steep density and/or rotation rate variations inside the star.

When a star fuses hydrogen in a thick shell around a core, consisting largely of helium, is subject to expansion. The "sub-giant star" cools and inflates. The rotation rate and the density alike increase in the core region and drop outward of the hydrogen burning shell. Mass is systematically transferred outwards and the process of angular momentum redistribution steadily reduces the rotational velocity for all the expanded outer layers. The rotation period becomes longer and longer. Therefore, there is a distinct shell wherein the rotation rate and the plasma density gradients become highly intensified. Consequently, the Torus structure can be readily formed in this shell producing a magnetized star.

A star ascending the red-giant branch (RGB) continues to produce more and more helium; the core of the RGB star increases its rotation rate, mass and temperature. The star is fully convective and the "first dredge-up" process takes place; fusion products are brought to the surface. That is, for a star in the RGB track, the rotation rate outward of the burning shell decreases (through mass transfer) by both, the star expansion and the dredge-up processes. Again, there is a certain shell wherein the density and the rotation rate gradients abruptly decrease and a Torus might be formed. Furthermore, a similar distinct shell obviously exists during the asymptotic giant branch (AGB) phase, the "second dredge-up" process and the next "dredge-ups" following thermal pulses. Consequently, the star will probably preserve its own magnetic field throughout all the transition phases from a main sequence star to a red giant (e.g., Arcturus) or a supergiant star (e.g., like Betelgeuse, a much more massive star).

Consequently, the dynamo action for a fully convective M-type star or a red giant or a supergiant is essentially the same; the scales are different. A displacement of the density and/or the rotation rate gradients will move the Torus structure and the curve $u_e=c'$ inwards or outwards, depending on the relation between $u_e$ and $c'$.

# 4. Discussion

## 4.1. Origin of the toroidal component magnetic field in solar-type stars

In general, we argue that the formation of a single Torus is essentially associated with poloidal and axisymmetric magnetic field topology, whereas the generation of a more complex topology, where the toroidal component prevails, can result from a two-Torus structure. However, the latter must be further clarified. Thus, we discuss this issue using the most representative solar paradigm, which typically shows exo-photosphere magnetic field with prevailing toroidal component. Since the toroidal electric current, flowing inside the Torus, can easily develop a poloidal magnetic field, the generation of the toroidal magnetic field, at first sight, is unanticipated.

The surface rotational speed over the solar equatorial plane, being ~2000 ms$^{-1}$, corresponds to the rotation rate of ~450 nHz. In the surface layer, from ~0.95 to 1 R$_\odot$, the differential rotation corresponds to a transition from ~15 nHz to zero or a rotation speed variation of ~66 ms$^{-1}$. That is, an observer on the surface sees subsurface charges with gradually increasing rotational speed from zero to ~66 ms$^{-1}$. Moreover, the same charge adjacent to the surface is subject to an additional northward motion due to the solar meridional circulation with a speed ~20 ms$^{-1}$. Eventually, the observer on the surface realises that ***the skin toroidal current is largely switched northwards***. Therefore, the mainly meridional electric current can readily produce the observed photosphere toroidal magnetic field potentially related to active sunspot pairs. We can suppose that the current is switched poleward at ~0.015 R$_\odot$ or ~10 Mm beneath the solar surface and at latitudes where the major magnetic activity is manifested. In this way, the two toroidal currents (flowing along the two Torus structures) can eventually build up the prevailing toroidal magnetic fields evident out of the surface.

Certainly, the more realistic scenario for the surface electric current is that it will deviate ~11$^{o}$ from the local meridional circle, at latitudes ±30$^{o}$. The latter may result from Joy's law (e.g., Dasi-Espuig et al., 2010) which determines that the magnetic axis for sunspot pairs is slightly inclined to the solar east-west direction, tilting from 3º near the equator to 11º at latitudes ±30º. In a sunspot pair, the leading sunspot is closer to the equator than the following one. Thus the ratio of the poloidal to the toroidal component will be greater than 0.2.

## 4.2. The ultimate dynamo action for fully convective M-type stars

The first spectropolarimetric observations of V374 Peg concerning a fully convective M dwarf (Donati et al. 2006; Morin et al. 2008) have revealed that these objects can host magnetic fields containing a long-lived strong dipolar component almost aligned with the rotation axis. Following the study of V374 Peg, a significant study including **_16 fully-convective stars_** with 0.15-0.35 $M_\odot$ and periods 5-10 d was initiated (Morin et al., 2010). They verified that two radically different types of ***large-scale magnetic fields*** are observed, either a weaker multipolar, non-axisymmetric field configuration in rapid evolution or a strong and steady axial dipole field, whereas no distinction between these two groups of stars can be made on the basis of mass and rotation only (Morin et al., 2011). They estimated that all the stars in the strong dipole regime have values comprised between 0.5-1.6 kG, whereas the values for those in the weak, multipolar regime are lower than 0.2 kG. Unavoidably, they discussed the conjecture of two different dynamo regimes, a weak- and a strong-field branch, and whether these different dynamo solutions could coexist over some range of parameters. Furthermore, they emphasized that the weak-field versus strong-field dynamo bi-stability is a promising framework to explain the coexistence of two different types of large scale magnetism in very low mass stars.

In the context of our work, there is essentially one and only one mechanism generating the observed quantitative and qualitative different magnetic field features. Our suggestion is that the well-established morphological features essentially result from the formation of a single or a two-Torus structure in the star's interior. The build-up of a single Torus leads to a well-organized dipole-like magnetic field topology; in contrast, the formation of two toroidal structures is accompanied by a weak, multipolar and solar-type magnetic field configuration.

The co-existence of a bimodal dynamo action in the category of fully convective stars implies that these stars have the possibility to generate dipolar or multi-polar, weak or strong, solar-type or not magnetic fields, although lacking the tachocline layer and the radiative region. Thus, one can hypothesize that an ultimate and unique dynamo action mechanism does not demand the tachocline, and therefore, the associated αΩ dynamo model may be not at work. Besides, if the same ultimate dynamo action takes place in stars with radiative envelope, then this hypothesis would dictate that the fundamental dynamo action is actually independent of the convection region, too. ***In this case, both the α² and the αΩ dynamo models are nothing more than fake hypotheses***; the magnetic activity may be due to an innovative concept related to an unusual behaviour of the material world.

Certainly, one would argue that, in principle, there are at least five radically different dynamo mechanisms, as follows: Two distinct mechanisms for the fully convective M type stars, two more mechanisms for the stars with radiative envelopes (i.e., one for the peculiar Ap stars with strong dipolar magnetic field and a second with the weakly magnetized A stars) and one additional mechanism with the solar-type stars having both radiative and convective regions separated by a layer (the so-termed tachocline for the Sun). Such a possibility of different mechanisms is expressed, for instance, from Yang et al. (2017). On the contrary, as it is emphasized, we scrutinize the perspective that a single and unique dynamo action is the ultimate mechanism at work, in all the cases. Certainly, our suggested mechanism is far from every typical current approach; however, our methodology may give a much better approximation for the physical dynamo mechanism and pave the way toward a much more reliable and realistic treatment. The whole contemporary theoretical approach is axiomatically the following: The magnetic field, in a fluid-plasma with specific velocity field, can be generated, developed and finally recycled; and this evolutionary track takes place in stellar scales and in endless repetitive cycles. ***This dogmatic approach, without currents and electric fields***, gives the impression that a new electrodynamics is discovered! We have to seriously rethink whether the overall MHD approach has become, and especially at this time, rather a misleading methodology than a useful tool. It seems that the MHD approach is nothing more than a modern confusing structure without exodus, like the ancient Labyrinth. The majority in the scientific community continues to steadfastly believe that infinite microstructures of eddy currents in a fully convective star will produce a strong and well-organized bipolar magnetic field in successive activity cycles with repetitive reverses in polarity! One may suppose that the frustration of the whole scientific community is transformed into a pseudo-scientific hypothesis; as a matter of fact, the science in this topic seems to be based on a purely unrealistic belief. The modern effort rather resembles the Sisyphean activity of king Sisyphus in the ancient Greek mythology.

After all, it is imperative in every research effort, and it is legitimate for every scientist, to pursue a more unified theory. The impartial searcher of truth, without prejudice, must thoroughly scrutinize all the possible questions. It is stressed that we focus our efforts to attain, if possible, a unified stellar dynamo action being at work under extremely variable stellar parameters and properties of matter. This work is the third one of this author testing his approach in the light of contemporary observations. Specifically now, we are interested in red stars within and off the main sequence, in the H-R diagram; whether the same mechanism plays a role in other type stars, for instance compact stars and magnetars, is an open question for future research.

It was postulated that a "turbulent dynamo" produces magnetic fields by random convective motions in the convection zone, and does not require rotation (or differential rotation) or a radiative-convective boundary layer for its operation. Thus, it was considered as a promising candidate for the dynamo action in late M-type stars. However, many initial key expectations or predictions for this model are not observationally confirmed; on the contrary, recent research shows something entirely different. In a book published about 20 years ago (with title "New Light on Dark Stars; Red Dwarfs, Low-Mass Stars, Brown Dwarfs", by Reid and Hawley, 2000), the two authors underlined that "the observational predictions of the turbulent dynamo are: (a) there is a weaker (or no) dependence on rotation; (b) there is no evidence for cyclic behaviour; and (c) a uniform coverage of active regions over the surface, rather than concentrated at mid-low latitudes is anticipated". Today none of these predictions is validated; in contrast, we have the opposite observational evidence, as we briefly discuss bellow.

Relatively to the issue of uniform coverage of active regions over the surface, Barnes et al. (2015) presented high resolution Doppler images of the M4.5 dwarf, GJ 791.2A, and the M9 dwarf, LP 944-20. The time series spectra of both objects revealed more starspot structures at high latitudes. The image of GJ 791.2A shows starspots located at a range of latitudes and longitudes, but preferentially at mid to high latitudes, whereas activity is confined solely to high latitudes on LP 944-20. Their images showed a number of polar or circumpolar spots, especially in GJ 791.2A. The concentration of spots seen on GJ 791.2A at 75°appear to occur in a relatively narrow band.

Furthermore, there is strong evidence for activity cycles in a slowly rotating, fully convective M5.5 star, with a length of approximately 7 years: the Proxima Centauri star (Mascareño et al. 2016; Wargelin et al. 2017). Actually, this star is a frequently flaring one with a rotation period of about 83 d (Kiraga and Stepien 2007; Mascareño et al. 2016). A dynamo simulation (Yadav et al. 2016), designed to mimic some of the physical characteristics of Proxima Centauri, was claimed to reproduce differential rotation in the convection zone that drives coherent magnetic cycles where the axisymmetric magnetic field repeatedly changes polarity at all latitudes with an activity cycle about nine years. Lavail et al., (2018) have identified that AD Leo (i.e., an extensively studied active M3Ve dwarf at the threshold of the fully convective regime), with dipole-dominated axisymmetric field topology, can undergo a long-term global magnetic variation.

In general, a relationship between the stellar rotation period and the stellar activity (which is commonly referred to as a "rotation–activity" one) has been observational revealed. Rapidly rotating cool stars are clearly more active than their slowly rotating counterparts. Moreover, many studies have suggested that the fully convective M-stars, with slow rotation periods, also follow a rotation–activity relationship similar to the stars with radiative core, i.e., a saturated activity below a threshold *Rossby* number Ro and a gradual decline for higher Ro (Kiraga and Stepien 2007; Reiners et al. 2009; Jeffries et al. 2011; Astudillo-Defru et al. 2016; Newton et al. 2017; Stelzer et al. 2016; Wright and Drake 2016). Therefore, even the fully convective stars probably develop a rotation–activity relationship, contrary to what was initially anticipated.

## 4.3. Comment on the work of West et al. (2015)

The conclusions resulting from the observational work of West et al. (2015) are explained very well in the context of this work. They used observations from 238 nearby M dwarfs for examining the rotation and magnetic activity of mid-to-late M dwarfs. We focus on a few findings which are summarized as follows: (1) For early-type (M1-M4) dwarfs, all stars rotating faster than 26 days are magnetically active and the strength of magnetic activity appears to decline with increased rotation period. (2) The late-type (M5-M8) dwarfs with rotation periods shorter than 86 days are also active and remain at a similar activity level as rotation slows, perhaps until the slowest rotation periods (> 90 days). (3) At the longest rotation periods, we see no active early-type M dwarfs, but there are a few slowly rotating, active, late-type M dwarfs. What is responsible for this mismatch?

In our view, the interpretation for the quoted conclusions is as follows: Obviously, the activity of a star, being tightly associated with its magnetic field, is not exclusively depended on its rotation speed. Apparently, the activity only for the early M dwarfs is directly related to the rotation period (i.e., there is a clear decrease in the strength of magnetic activity with increasing rotation period). The reason is that, for the fully convective stars, a similar decrease is counterbalanced from an antagonistic action: The particularly intensified shear layer L1 (described in subsection 2.1 and resulting from the gradient of the rotation rate) becomes particularly important in this situation. Consequently, at the longest rotation periods (i.e., from T=26 to 86 d) there are active stars exclusively within the category of late M dwarfs. ***For these stars the role of the rotation speed is minimized; in contrast, the extremely intense gradient of the shear layer L1 is recognized being the dominant factor***. Both, the rotation speed and the $\nabla\Omega$ alike, increase the total amount of the toroidal current flowing within the Torus. Based on this suggestion, one must anticipate an intense increase of magnetic activity occurring at about the M4-type stars. And the latter is discussed in the following subsection.

## 4.4. Preferentially increased occurrence rate of superflares in red dwarfs

The fraction of M spectral type stars that are determined as active, on the basis of the Hα emission line as an indicator, peaks at spectral type M8, where 73% of stars are active. In this regard, (West et al., 2004). Yang et al., (2017), carrying out a statistical study of flares using 540 M dwarfs from the Kepler Space Observatory have reached the conclusion that the flare activity and the number fraction of flaring stars in M dwarfs rise steeply near M4. A few years ago, Candelaresi et al., (2014) observed a clear decrease of the superflare frequency with increasing effective temperature using *Kepler* data from 1690 superflares (with energy higher than $10^{34}$ erg) corresponding to G-, K-, and M-type stars. Such valuable results are interpreted in our model as follows: We know that the lower the effective temperature (for a main-sequence-star), the higher will be the mean star's density (Zombeck, 2007). We further know that the higher mean density is associated with an increased occurrence rate of superflares. The reason of this behaviour is related, in our dynamo action, to the assumption that the tachocline is not particularly important; the tachocline may exist (in stars with convective envelopes) or not (in fully convective stars). What is so important? We identify the critical role on a layer at which the radial density abruptly drops. Such a layer with steep density gradient (pointing inward) separates the star's outer region characterized by plasma meridional circulation from the rest of the star. In this outer portion of the star the rotation rate increases because the angular momentum is redistributed. Therefore, the layer related to an abrupt density drop is actually associated with the differential rotation of the star. Within this shear layer one or two Torus structures might be potentially generated and developed. Moreover, the larger the mean density, the closer to the photosphere the Torus structures are proportionally formed, leading to more intense superflares characterized by higher occurrence rates, too.

The adoption of the above proposed scenario can explain the occurrence of very extreme events like those called "mega flares"; for instance, a representative one occurred on April 23, 2014 (Fender et al., 2015). At that time the strongest, hottest, and long-lasting sequence of stellar flares ever seen occurred. The mega flare is produced by a red dwarf star of spectrum type M4.0Ve belonging to the binary star system DG CVn. Certainly, at least one of the star system components is a very rapid rotator with vsini≈50 kms$^{-1}$; thus one may argue that exclusively the rotation speed is the unique factor of activity. However, superflares are observed at slow rotating stars, too; obviously, the activity is depended on additional parameters.

Proxima Centauri, being of spectral type M6Ve, is a fully convective star with about one-seventh the actual diameter of the Sun and average density about 33 times that of the Sun. Most importantly, Proxima is characterized by a very low rotational speed: From photometric observations, Benedict et al. (1998) found rotation period $P_{rot}$=83.5 days and rotational velocity (v sin i) less than 100 m/s. Although Proxima has a very low average luminosity; however, it is a flare star that undergoes random dramatic increases in brightness because of magnetic activity. In March 2016, the *Evryscope* observed the first "naked-eye-brightness superflare" detected from Proxima. The superflare had a bolometric energy of $10^{33.5}$ erg, while it was predicted that at least five superflares occur each year (Howard et al., 2018).

West and Basri (2009) stressed that their results suggest that rotation and activity in late-type M dwarfs may not always be linked together. For us the additional parameters are (1) the systematically increasing mean star density and (2) the outward mass transfer in the fully convective stars. The latter produces steep gradients of the rotation rate in the shear layer related to the Torus structures.

# 4.5. Four key parameters associated with the magnetic field of red stars

In general, *we identify four key parameters* associated with the magnetic field of red dwarfs and red giants and supergiants: **First**, the rotation speed; **second**, the steepness of the radial gradient of the rotation rate ($\nabla\Omega$) in the shear layer; **third**, the distance of the Torus from the photosphere and **fourth**, the cross-sectional area of the Torus. In our framework, we could anticipate that (a) the higher the rotation speed, the greater the toroidal current flowing inside the Torus; (b) the steeper the gradient of the rotation rate, the greater the electron acceleration or deceleration generating a larger amount of positive or negative charge, respectively, within the shear layer; (c) the deeper from the surface the Torus is formed, the weaker the detected photosphere magnetic field and (d) the greater the Torus cross-sectional area, the greater the toroidal current. The first two parameters are included in past research efforts and in this work, too. We have underlined that the steepness of the shear layer is an absolutely pre-required feature for generating magnetic field. Similarly, the relatively higher rotation speed always produces stronger field. However, *the Torus distance from the photosphere, as a major and independent parameter, was not introduced in the past; nevertheless, this parameter essentially dictates the scale factor of the magnetic field for red stars*. That is, the macro-scale level of the magnetic field is determined by this amount. The first two parameters contribute by increasing the Torus current density, J; that is, the rotation velocity, $u_{rot}$, and the volume charge density, nq, increase the J according to the relation $J=nqu_{rot}$. In turn, the Torus current gives rise to strong poloidal magnetic fields that form the surface/subsurface charged layer directly related to the surface magnetic field. The importance for the third key parameter is stressed in the next two subsections, while the fourth parameter is introduced later on, in subsection 4.10.

## 4.6. Our dynamo model related to a past statistics of 48 red giants

Aurière et al., (2015) detected magnetic field via Zeeman signatures in 29 out of 48 red giants. This is the best presently available statistics and their results are well interpreted in the context of our dynamo model. Thus, it is meaningful to look at their datasets in the light of this research effort. The importance for the first three key parameters (cited in the preceded subsection) becomes apparent. We underline a few aspects:

1. First from the 29 red giants with measured magnetic field, we select all the stars **with radius greater than 25 $R_\odot$**. The six stars (i.e., nu3 Cma, 37 Com, η Psc, Aldebaran, Alphard and Arcturus) in this class, give the average values **<B>=1.67 G** and **<$R_*$>= 39.87 $R_\odot$**, respectively. In contrast, if we take all the stars **with radius less than 8 $R_\odot$**, then we pick up 7 stars (i.e., 14 Cet, EK Eri, V390 Aur, FI Cnc, 24 Uma, δ CrB, and OU And) with **<B>=31.1 G** and **<$R_*$>= 5.8 $R_\odot$**, respectively. As a matter of fact, *<u>the larger the $R_*$ value is, the weaker the magnetic field becomes</u>*. We infer that the Torus is formed deeper from the surface, when the star is bigger.

2. Second, we focus on the three stars (i.e., EK Eri, 14 Cet and OU And) which are termed "outliers" in their statistics, because of their extremely strong magnetic field levels. Their magnetic fields and radii have the average values **<B>=58.7 G** and **<$R_*$>= 5 $R_\odot$**, respectively. In conclusion, *the strongest magnetic fields are associated with the lowest values of $R_*$*. Consequently, we infer that the star's radius plays a decisive role and according to our interpretation, for these three stars, the Torus is formed very close to the surface, and perhaps the shear layer is steeper.

3. Third, we select the red giants having vsini greater than 20 kms$^{-1}$; that is, we select the stars with the greatest rotational speeds. In result, we have a category of five stars (i.e., V1192 Ori, V390 Aur, 31 Com, KU Peg and OU And) well above the sub-Gauss class. Their mean values of vsini and the longitudinal magnetic field are **<vsini>=35.9 kms$^{-1}$** and **<B>=18.2 G**, respectively. Conversely, if we select all the stars with vsini less than 5 kms$^{-1}$; then we have a category of 11 stars (i.e., EK Eri, 77 Tau, 19 Pup, 39 Hya, η Her, ξ Her, ρ Cyg, ε Tau, Aldebaran, Pollux and Arcturus) with **<vsini >=3.3 kms$^{-1}$ and <B>=12.2 G**, respectively. Certainly, we have to stress that the latter statistics is largely affected by the EK Eri star with 98.6 G; without it **<vsini >=3.6 kms$^{-1}$** and **<B>=3.5 G**. Actually, *within the same class of giant stars, the higher the rotational speed, the stronger the magnetic field*.

Our inference that the magnetic field magnitude (for red giants) is basically controlled from the size of the star's radius is based on stars having much lower and much greater radii. In this way, one based on the star's radius as a dominant parameter can hypothesize that the late M-type red dwarfs must be related to much stronger magnetic fields, whereas the red supergiants must develop extremely weak magnetic fields. Actually the latter is observationally confirmed; the related magnetic fields range from hundreds of Gauss to ~1 Gauss; and this topic is particularly elaborated in the next subsection.

# 4.7. Magnetic field magnitude for red dwarfs, red giants and red supergiants controlled by the star's radius

We bellow mention three statistical works corresponding to the magnetic field observations for red dwarfs, red giants and red supergiants, respectively. A significant inference is derived:

1. Morin et al., (2010) examined *11 fully convective M dwarfs* (of spectral type M5–M8) and observed a first category of five stars (i.e., GJ51, GJ1154A, GJ1224, CNLeo and WXUMa) with strong dipole-like magnetic field and a second category of six stars (i.e., GJ1156, GJ1245B, DXCnc, GJ3622, VB8 and VB10) with weak and complex field. The average values of the longitudinal magnetic fields are <B>=978.3 G and <B>=48.2 G in each category, while the strongest observed field is 1642 G for the star WXUMa. Finally, the magnetic field and the radius (for all the 11 stars) is ranged as follows: **29<B<1642 G** and **0.09<R$_*$<0.22 R$_\odot$.**

2. Auriére et al., (2015) detected magnetic fields for *29 red giant stars*. For these stars the maximum values of the longitudinal magnetic fields and their radii are ranged as follows: **0.34<B<98.6 G** and **2.8<R$_*$<66.1 R$_\odot$.**

3. Tessore et al., (2017) searched for the surface magnetic field in a sample of a few late-type *red supergiant (RSG) stars,* being very massive-cool evolved stars. For the stars CE Tau, α Ori and μ Cep, the longitudinal component of the detected surface fields is at the Gauss-level, while the field of α$^1$ Her is an order of magnitude stronger (i.e., 6 G). Additionally, they reported that in the yellow supergiant star ρ Cas magnetic field is not detected. In conclusion, all these four magnetic stars with **1<B<6 G** are fully convective (with the star α$^1$ Her on the "asymptotic giant branch") with a common feature: The radius is roughly extended in the range **300<R$_*$<1500 R$_\odot$.**

In conclusion, if the subscripts RSG, RG and RD denote red supergiants, red giants and red dwarfs, respectively, then we roughly could write the following relations:

$$R_{RSG} \approx 50\ R_{RG} \text{ and } R_{RG} \approx 50\ R_{RD}$$

$$B_{RSG} \approx B_{RG}/15 \text{ and } B_{RG} \approx B_{RD}/15$$

$$M_{RSG} \approx 5\ M_{RG} \text{ and } M_{RG} \approx 10\ M_{RD}$$

Obviously, the extremely low densities in giants, and particularly in supergiants, indicate that the curve $u_e = c'$ is probably displaced more inward. ***The larger the star, the deeper inward the Torus is formed.*** **Figure 3** schematically shows the supposed position of the curve $u_e = c'$ for the three classes of stars; that is, red dwarfs, red giants and red supergiants, in r/R$_*$. Therefore, ***the strength of the magnetic field, for stars of the same spectral type, is essentially controlled by the size of the star.***

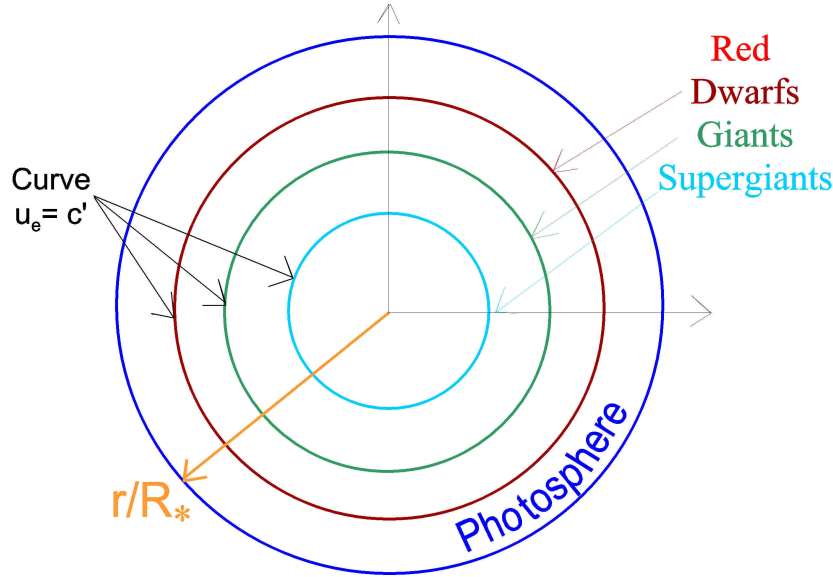

***Fig. 3.*** *Sketch showing the supposed position of the curve $u_e = c'$, in r/R$_*$, for red dwarfs, red giants and red supergiants.* ***Stars of the same spectral type form*** *the Torus structure*

*gradually deeper. The distance of Torus from the photosphere determines the order of magnitude of the magnetic field.*

## 4.8. The case of the EK Eri-star

EK Eri (with 4.7 $R_\odot$ and 2 $M_\odot$) is the most known as a slowly rotating active G8-giant; however, contrary to its long rotation period (309 d), the magnetic field is ~100 G, which could not be maintained by a classical, solar-type, dynamo (Aurière et al., 2008 and references therein). Thus, it was proposed that EK Eri is the descendant of a strongly magnetic Ap star. It was argued that, at its current evolutionary phase, the star experiences the so-called first dredge-up, while the convective envelope of the 2 $M_\odot$ model contains 0.37 $M_\odot$ (Aurière et al., 2011). Obviously, in the case of the EK Eri-star, strong and well-structured magnetic field co-exists with a very low rotational speed; a contradictory synthesis. In the context of our work, the above antinomy could be resolved as follows: We have already underlined that the rotation velocity is neither the unique, nor always the dominant parameter. Moreover, one can claim that, perhaps, the role of the rotation speed was overemphasized in the past. Conversely, a steep reduction of the rotation rate or a steep radial gradient of plasma density (in the star's interior) may suffice producing the observed magnetic field; such a possibility should be anticipated. Therefore, we suggest that sometimes the rotation speed can't be considered as the only decisive factor generating magnetic field. A layer, characterized by an abrupt density or rotation rate change, would accumulate a huge amount of charge. Consequently, an exceptionally intense toroidal current would be built up, flowing along the Torus. The star would be strongly magnetized despite its own extremely low rotational speed (of ~300 ms$^{-1}$). A single Torus suffices producing the large-scale strong and poloidal magnetic field. The adoption of the statement "an Ap star is the progenitor of EK Eri" rather covers up the real problem of dynamo action; the question is essentially shifted in a preceded phase.

## 4.9. Dynamo theory (α-Ω) questioned

The study on the relationship between coronal X-ray emission and stellar rotation in late-type main-sequence stars confirms that two emission regimes exist, one in which the rotation period is a good predictor of the total X-ray luminosity at high rotation periods, and the other in which a constant saturated X-ray to bolometric luminosity ratio, $L_x/L_{bol} \approx 10^{-3}$, is attained at low rotation periods (e.g., Pizzolato et al., 2003). Moreover, it is shown that the X-ray emission saturation occurs below a critical rotation period, $P_{sat}$, increasing with decreasing stellar mass. Wright and Drake, (2016) reported observations of four fully convective stars whose X-ray emission correlates with their rotation periods in the same way as in Sun-like stars. Furthermore, Wright et al., (2018) confirmed the existence of fully convective stars in the X-ray unsaturated regime and found that these objects follow the same rotation–activity relationship seen for partly convective stars. They stressed that these results show that fully convective stars, at least when they have spun down sufficiently, operate a dynamo that exhibits a rotation–activity relationship that is indistinguishable from that of solar-type stars. Their conclusion is that fully convective stars also operate a solar-type dynamo in which the tachocline is not a critical ingredient, given that the X-ray activity–rotation relationship is a well-established proxy for the behaviour of the magnetic dynamo. This conclusion, in the context of this work, is supported via a completely different dynamo action mechanism. Our dynamo, which was initially proposed at 2017 and improved at 2019, is potentially at work for fully convective red dwarfs, red giants and supergiants, too. It is mainly based on the extremely strong levels of radial shear just beneath the surface and the formation of the powerful Torus structure. ***The tachocline, as an interface layer between the radiative core and the convective envelope, does not play a significant role in the generation of the magnetic field.*** Therefore, we line up with Charbonneau (2016) that the whole dynamo theory (i.e., the α-Ω dynamo action) should be questioned.

# 4.10. The reason for the saturation in the rotation–activity relationship

For very fast rotators, the rotation–activity relationship has been found to break down, with X-ray luminosity reaching a saturation level of approximately $L_X/L_{bol} \approx 10^{-3}$, independent of the stellar spectral type (e.g., Vilhu, 1984; Wright et al., 2011). We shall attempt an explanation for this inference in the context of our dynamo model.

In a first place, we assume that (a) the Torus toroidal electric current is proportional to its cross sectional area and (b) the volume charge density of Torus, as well as its rotational speed, remain unvaried during the charging process. In a second place, the stellar magnetic activity is assumed being directly related to two factors: The level of the Torus electric current and the depth beneath the surface in which the Torus is formed. According to the preceded subsections, the smaller the star's radius, the closer to the surface the Torus is formed. Thus, we focus on the Torus current density, which is proportional to its volume charge density, nq, and the rotational speed $u_{rot}$ (or the rotational period); that is, $J=nqu_{rot}$. And obviously, the total current is directly depended on the Torus cross-sectional area, $\pi r^2$. Moreover, the volume charge density, which is tightly related to the steepness of the radial shear, and the rotational speed of the Torus determine the current density. Once the Torus is established, its poloidal magnetic field attracts more and more charges inside; as a matter of fact ***the Torus cross-sectional area steadily increases. However, the Torus radius enhancement has an upper limit***; the limit is set by the curve $u_e=c'$. At greater radii the Torus intersects the curve $u_e=c'$ and loses its exotic property accumulating additional charge. In effect, the Torus current remains unvaried and ***the magnetic activity reaches its saturation level***. That is, the combined action of both, the rotational speed and the level of radial shear, results in magnetic saturation. One can estimate the volume charge density of Torus assuming that the electric force equals the magnetic force and that the latter is achieved at a scale length $L=\kappa\lambda_D$, where $\lambda_D$ is the Debye length.

The same level of X-ray to bolometric luminosity ratio is achieved in fully convective M dwarfs with longer periods as compared to partly convective stars (Wright et al., 2018; look at their figure 2). In our approach this is mainly due to the fact that in a star with smaller radius the Torus is formed closer to the surface; thus, a given activity level could be achieved with longer periods.

# 5. Conclusions

Major conclusions:

1. In general, the magnetic field observations for red dwarfs, red giants and red supergiants can be reproduced by the Torus structure (as introduced by Sarafopoulos, 2017, 2019), in cooperation with additional more specific processes like the differential rotation, mass transfer, angular momentum redistribution, rotational speed etc. Eventually, ***all our three works propose a promising unified dynamo action, being potentially at work for main sequence stars as well as for red giants and supergiants.***

2. For red dwarfs, we suggest that the formation of a single Torus is essentially associated with a large-scale strong, poloidal and axisymmetric magnetic field topology, whereas the generation of a weaker multipolar, non-axisymmetric field configuration in rapid evolution can result from a two-Torus structure (like the solar case).

3. ***We identify four key parameters*** associated with the magnetic field of red dwarfs, giants and supergiants: **First**, the rotation speed; **second**, the steepness of the radial gradient of the rotation rate ($\nabla\Omega$); **third**, the distance of the Torus from the photosphere and **fourth**, the cross-sectional area of the Torus. The third and fourth key parameters are introduced for the first time.

4. Both, the rotation speed and the $\nabla\Omega$ alike, increase the total amount of the toroidal current flowing within the Torus. Based on this principle, one must anticipate (a) an intense increase of magnetic activity occurring at about the M4-type stars, and (b) an increased activity for only the fully convective M-type stars. Actually, both expectations are supported by measurements.

5. The strength of the magnetic field, for stars of the same spectral type (i.e., red dwarfs-RD, red giants-RG and red supergiants-RSG), is essentially controlled by the size of the star. We suggest that, ***the larger the star, the deeper inward the Torus is formed***. The Torus distance from the photosphere essentially dictates the scale factor of the magnetic field. Actually, we can reveal the following relations from already published observations:

$$R_{RSG} \approx 50\ R_{RG} \text{ and } R_{RG} \approx 50\ R_{RD}$$
$$B_{RSG} \approx B_{RG}/15 \text{ and } B_{RG} \approx B_{RD}/15$$

$$M_{RSG} \approx 5\ M_{RG} \text{ and } M_{RG} \approx 10\ M_{RD}$$

6. After the Torus generation, its poloidal magnetic field attracts more and more charges inside; in effect, its cross-sectional area steadily increases. However, the Torus radius enhancement has an upper limit; at greater radii the Torus intersects the curve $u_e = c'$ and loses its exotic property of accumulating additional charges. Thus, the Torus current remains unvaried and ***the magnetic activity reaches its saturation level***, as it is defined by the rotation-activity relationship.

7. On the basis of our proposed dynamo action, being potentially at work in fully and partly convecting stars, as well as in stars with radiative envelopes, the α-Ω and $α^2$ dynamo models become non-realistic. In effect, the tachocline, as an interface layer between radiative core and convective envelope, does not play its supposed role in the α-Ω model.

8. We explain why the toroidal component of the photosphere magnetic field prevails in the solar-type dynamo action paradigm. Initially, the Torus triggers an azimuthal electric current flowing in a thin subsurface layer characterized by differential rotation. In turn, this current flow is switched poleward by the established meridional circulation of the solar plasma. Thus, at latitudes where particularly increased activity is observed, the magnetic field is mainly toroidal.